\documentclass[showpacs,twocolumn,prb,eqsecnumbers]{revtex4}
\usepackage{amsmath}
\usepackage{epsfig,color}

\usepackage{graphicx}
\usepackage{dcolumn}
\usepackage{bm}

\def\mb{\mathbf}

\newcommand{\eps}{\varepsilon}

\newcommand{\beq}{\begin{equation}}
\newcommand{\be}{\begin{equation}}
\newcommand{\eeq}{\end{equation}}
\newcommand{\ee}{\end{equation}}
\newcommand{\bea}{\begin{eqnarray}}
\newcommand{\eea}{\end{eqnarray}}
\newcommand{\bwt}{\begin{widetext}}
\newcommand{\ewt}{\end{widetext}}

\begin{document}

\title{Non-Landau damping of magnetic excitations in systems with localized and itinerant electrons}
\author{Andrey V. Chubukov$^{1}$, Joseph J. Betouras$^{2}$ and Dmitry V. Efremov$^{3}$}
 \date{\today}

\begin{abstract}
We discuss the form of the damping of magnetic excitations in
a metal near a ferromagnetic instability.
The paramagnon theory predicts that the damping term should have the form $\gamma(q,\Omega) \propto \Omega/\Gamma (q)$ with $\Gamma (q) \propto q$ (the Landau damping).
However, the experiments on uranium metallic compounds UGe$_2$ and UCoGe showed that $\Gamma (q)$
 is essentially independent on $q$.
A non-zero
 $\gamma(q=0, \Omega)$
is impossible in systems with one type of carriers (either localized or itinerant) because it would violate the spin conservation.
It has been conjectured recently  that a
near-constant $\Gamma (q)$ in UGe$_2$ and UCoGe
may be due to the presence of both localized and itinerant electrons in these materials, with ferromagnetism involving predominantly localized spins.  We present the microscopic analysis of the damping of near-critical localized excitations due to
interaction with  itinerant carriers.
We show explicitly how the presence of two types of electrons breaks the cancellation between the contributions to $\Gamma (0)$ from the
self-energy and vertex correction insertions into the spin polarization bubble.
We compare our theory with  the available experimental data.
\end{abstract}

\affiliation{
$^{1}$
 Department of Physics, University of Wisconsin-Madison, 1150 University
Ave., Madison, WI 53706-1390, USA.
\\
$^{2}$ Department of Physics, Loughborough University,Loughborough LE11 3TU, UK. \\
$^{3}$ Leibniz-Institute for Solid State and Materials Research, IFW-Dresden, D-01171 Dresden, Germany.}

\maketitle

{\it Introduction.}---
Recent progress in neutron scattering measurements of spin structure factor in uranium metallic materials
UGe$_2$ and UCoGe provided detailed
information on the dynamical structure factor of a paramagnetic metal
near the transition into an itinerant ferromagnet~\cite{Huxley2003,Raymond2004,Stock2011}.
The dynamical structure factor $S(\mb{q},\Omega)$ at wave-vector $\mb{q}$ and frequency $\Omega$
is related to the dynamic spin susceptibility $\chi (\mb{q}, \Omega)$ via the relation
$S(\mb{q},\Omega) =  \chi^{''} (\mb{q},\Omega) \coth \frac{\Omega}{2T}$.
In a conventional paramagnon theory of a nearly ferromagnetic metal
$\chi (\mb{q},\Omega) \propto 1/(1 - i \Omega/\Gamma (q))$
and $S(\mb{q},\Omega) \propto \Gamma(\mb{q})/(\Omega^2+ \Gamma (q)^2)$ at $\Omega \ll v_F q$,
with $\Gamma (q) \propto q$ for clean systems and  $\Gamma(q) \propto q^2$ in the presence of
 non-magnetic impurities \cite{Halperin1969}.

This behavior has been
observed in three-dimensional (3D) electron itinerant ferromagnets such as Ni, Ni$_3$Al, Fe and MnP \cite{NiandFe, Ni3Al, MnP}.
Experiments on UGe$_2$ and UCoGe, however, detected a different behavior -- near $T_c$, $\Gamma (q)$ extrapolates to a finite value at $q\to 0$
both in the paramagnetic and the ferromagnetic state.

In
one-component fermionic systems
the vanishing of
$\chi''(q\to0, \omega)$ is the consequence of the conservation of the total fermionic spin $\mb{S}_t$.
In particular, 
 $\chi''(q=0,\Omega) = 0 $ holds
  when the fermion-fermion interaction
is mediated by their own collective spin excitations.
 The situation changes, however, if there are several different bands crossing the Fermi level
 ~\cite{Mineev2012}, or magnetic impurities
 and/or strong spin-orbit coupling~\cite{cm_92},
 or if
 the system is a two-component one and the neutron scattering measures predominantly the contribution from only one of them in some range of $q$.

 This last case is
  applicable to UGe$_2$ and UCoGe because these systems possess simultaneously itinerant and localized
   ( $5f^2$) electrons \cite{Yaouanc2002}.
     In a recent work~\cite{Mineev2013}, Mineev argued that the spin response at small but finite $q$
     is predominantly determined by localized spins, because they predominantly contribute to long-range magnetic order
      and to static susceptibility above the Curie temperature (the contribution from itinerant fermions to magnetic order is around $1\%$, Ref.\onlinecite{Yaouanc2002,relax}).
  He presented a phenomenological description of the relaxation of localized spins due to interaction with itinerant carriers
  and  obtained $\Gamma (q)$ which only weakly depends on $q$.
 Indeed, the total spin of the localized and itinerant fermions is a conserved quantity, and strictly at $q=0$ and finite $\Omega$, $S(0,\Omega)$ it must vanish, i.e.,
  the contributions to $S(0,\Omega)$ from localized and from itinerant fermions must cancel out.
  However, because the static susceptibility of the localized spins is much larger than the one for itinerant fermions, the contribution to $S(q,\Omega)$ from the localized spins
   dominates down to
   $v_F q  \ll \Omega$.
The experiments on UGe$_2$ are performed well inside this range, at $v_F q \geq \Omega$ (Ref.\onlinecite{comm_new}).  The total spin of the localized fermions
is not a conserved quantity, and the corresponding $\Gamma (q)$ does not have to vanish at $q\to 0$.

In the present work we address the same issue from a microscopic perspective.  We show that the analysis of $\Gamma(q)$ at $q \to 0$ for the systems with spin-spin interaction is rather non-trivial,
 as the calculation of a spin-polarization bubble for interacting fermions requires
 the consideration, on equal footing, of the renormalizations coming from (i) the self-energy diagram, (ii) the Maki-Thompson (MT)-type vertex correction diagram, {\it and} (iii) the Aslamazov-Larkin (AL)-type diagrams.
The latter have formally one extra power of the coupling and are seemingly less important, but, as we will see later, for one-component systems they are actually of the same order as the other
terms. The reason is that
the extra power of the coupling gets absorbed into the fermionic damping which contains the same coupling.
The importance of including the two-loop AL diagrams into the analysis of the spin susceptibility of a one-component system at $q \to 0$
has been emphasized in Ref. \onlinecite{cm_prl}, where the authors
considered a Fermi liquid with the self-energy
$\Sigma (\omega) = \lambda \omega$
and demonstrated
that the combination of self-energy,  MT, and AL corrections  preserves spin conservation.
Here we demonstrate this for the more general case, when the self-energy also includes thermal damping $\Sigma^{''} (T)$.
Further we point out  the difference between one- and two-component systems.
We argue that for two-component systems, the AL diagrams become irrelevant at weak coupling because the balance between the damping and coupling is lost and the extra power of the coupling is not
cancelled out.
The remaining self-energy and MT insertions into the spin polarization bubble add up
and yield a non-zero $\Gamma (q=0)$, which is proportional to
the imaginary part of the single-particle self-energy of a conduction electron $\Sigma^{''} (T)$.
We compute $\Sigma^{''} (T, \omega)$ due to interaction with near-critical localized carriers and show that
it has a peak at $T_c$, in agreement with the experiments~\cite{Stock2011}.

\begin{figure}[tbp]
\begin{center}
\epsfxsize=1.0\columnwidth
\epsffile{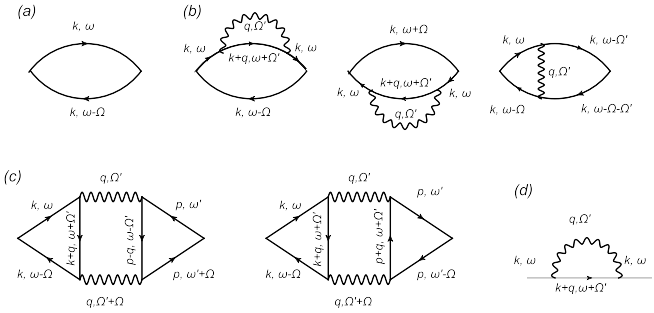}
\end{center}
\caption{ a-c) The spin polarization bubble $\chi (q, \Omega)$.  The thin lines are fermionic propagators, the wavy line is the effective interaction
 mediated by the total spin susceptibility $\chi^{tot} (q, \Omega)=\chi (q, \Omega)/(1-{\bar U} \chi (q, \Omega))$. Each side vertex and each interaction vertex contains spin $\sigma-$matrices (not specified). a) The bubble for free fermions $\chi_0 (q, \Omega)$; b) spin-polarization bubble with self-energy and MT-type vertex corrections
 c) The two AL diagrams. By power counting, they are of higher order, but in fact, they are of the same order as the diagrams in b) (see text).
 d) One-loop diagram for the self-energy of an itinerant fermion in a a two-component model, due to interaction with localized spins. The
  wavy lines are $\chi_L (q, \Omega)$ for
  localized electrons. }
 \label{fig1}
\end{figure}

{\it One-component itinerant system.}---
To set the stage for the analysis of the two-component systems, consider first a one-component system of itinerant electrons, for which the
the total electronic spin $\mb{S}_t$ is conserved.
The conservation of $\mb{S}_t$ implies that its zero-momentum Fourier component
does not depend on time $t$, hence both
$\chi ({\bf{q}},\Omega)$
and the dynamical structure factor
  $S({{\bf{q}}},\Omega)$
should vanish at $q=0$ and
any non-zero frequency $\Omega$.
As an example
(and to simplify the formulas),
  consider a two-dimensional ($2D$) system with isotropic dispersion
 $k^2/(2m)$.
For non-interacting fermions
 $\chi_0 ({\bf{q}},\Omega)  = (m/\pi) (1 - \Omega/ \sqrt{\Omega^2 - v^2_F q^2})$,
    and
     $\chi_0 (0,\Omega)$ and $S_0(0,\Omega)$ vanish, as they should.  At finite $q$ and at $\Omega \ll v_F q$,
$\chi_0 ({\bf{q}}, \Omega) \approx (m/\pi) (1 + i \Omega/\Gamma_0 (q))$
with  $\Gamma_0 (q) = v_F q$. This quantity is non-zero, however it vanishes at $q=0$.
In the presence of disorder, $\Gamma_0 (q)$ acquires a diffusive form $ \Gamma_0 (q) = \sqrt{v^2 q^2 + \gamma^2} - \gamma$, where
 $\gamma$ is the impurity scattering rate,
 and it still vanishes at $q=0$ ($\Gamma_0 (q) \propto q^2$ at small $q$).
We emphasize that the vanishing of $\Gamma_0 (q=0)$ directly follows from spin conservation, because if $\Gamma_0 (0)$ was finite, it would immediately imply a nonzero $S_0 (0, \Omega)$, in violation of the spin conservation.

Because the four-fermion interaction $U$ is $SU(2)$ spin invariant,  $\Gamma (q=0)$ should vanish in
 an interacting system as well.
 The total susceptibility $\chi^{tot} ({\bf q},\Omega)$ is  proportional to the fully renormalized spin polarization bubble $\chi ({\bf q},\Omega)$
(a fully dressed  particle-hole bubble with spin $\sigma-$ matrices in the vertices). In the Random Phase Approximation (RPA),
  $\chi^{tot} ({\bf q},\Omega) = \chi ({\bf q},\Omega)/(1-{\bar U} \chi ({\bf q},\Omega))$, where ${\bar U} = U/2$.
 Using the free-fermion form of $\chi ({\bf q},\Omega))$ one immediately reproduces the paramagnon formula
 $S({\bf{q}},\Omega) \propto \Gamma(q)/(\Omega^2+ \Gamma (q)^2)$, with $\Gamma (q) \propto q$.

The conservation of the total spin implies that the fully renormalized $\chi^{tot} ({\bf q},\Omega) \propto \chi ({\bf q},\Omega)$ should vanish at ${\bf q} =0$.
For free fermions, this holds, as we demonstrated above.
 To go beyond free fermions, we need a model for fermion-fermion interaction.
With the uranium compounds in mind, we consider a nearly ferromagnetic metal.
Following earlier works~\cite{early},
 we assume that the low-energy physics  is described by an effective model in
which the bare  $U$ is replaced by an effective dynamical interaction $U_{eff} (q, \omega)$ in the spin channel, which is mediated by
 $\chi^{tot} ({\bf q}, \omega)$.
 In the RPA, $U_{eff} (q, \omega) = {\bar U}^2 \chi^{tot} (q, \Omega) = {\bar U}^2  \chi ({\bf q},\Omega)/(1-{\bar U} \chi ({\bf q},\Omega))$.
 Using ${\bar U} \chi (0,0) \approx 1$, valid near a ferromagnetic transition, this can be simplified to $U_{eff} (q, \omega) \approx {\bar U}/(1 -{\bar U} \chi ({\bf q},\Omega)))$.
 We will
 show how the condition $\chi (q=0,\Omega) =0$ is satisfied at each order
 in ${\bar U}$.

The diagrams for the $\chi(q=0,\Omega)$ to order ${\bar U}$ are shown in Fig. 1b. The first two diagrams contain the self-energy insertions, the third is the MT vertex correction diagram.  Each of the three diagrams contains the product of four Green's function and one dynamical $U_{eff} (q, \omega)
  \propto {\bar U}$.
  In explicit form $\chi (q=0,\Omega) =  I_{se} (\Omega) + I_{MT} (\Omega)$
 is~\cite{see}:
 \begin{widetext}
  \bea
  && I_{se} (\Omega) = -6 \int \int G^2(\mb{k},\omega) G(\mb{k},+\mb{q}, \omega+ \Omega') \left(G(\mb{k}, \omega-\Omega) + G(\mb{k}, \omega +\Omega)\right) U_{eff} (\mb{q},\Omega') \nonumber \\
  && I_{MT} (\Omega) = 2  \int \int G(\mb{k},\omega) G(\mb{k}, \omega-\Omega) G(\mb{k},+\mb{q}, \omega+ \Omega') G(\mb{k}+\mb{q}, \omega-\Omega + \Omega') U_{eff} (\mb{q},\Omega')
 \label{1}
 \eea
 \end{widetext}
 where the prefactors are due to the summation over spin indices, $G (k,\omega) = (\omega - \eps_k + i \delta {\text sgn} \omega)^{-1}$,  and we adopt the notation $\int \int \equiv \int \frac{d^dk d\omega}{(2 \pi)^{d+1}}\int \frac{d^d q d\Omega'}{(2 \pi)^{d+1}}$
  for a $d$-dimensional system.
 Applying several times the identity:  $G(\mb{k}, \omega) G(\mb{k}, \omega -\Omega) = (G(\mb{k}, \omega-\Omega)- G(\mb{k}, \omega))/\Omega$,
 we explicitly re-write the two terms as
  \be
  I_{se} (\Omega) = 6  A (\Omega), I_{MT} (\Omega) = 2 A(\Omega)
   \label{2}
   \ee
   where
  \bea
   A(\Omega) = &&\frac{1}{\Omega^2} \int \int \left(2G(\mb{k},\omega) - G(\mb{k}, \omega-\Omega) -G(\mb{k}, \omega+\Omega)\right) \nonumber \\
    && \times G(\mb{k}+\mb{q},\omega + \Omega') U_{eff} (\mb{q},\Omega')
  \label{3}
  \eea
  We see that the self-energy and MT diagrams are of the same sign and add up:
   $I_{se} (\Omega) + I_{MT} (\Omega) = (4/3) I_{se} (\Omega)$.
  The frequency-dependent part of
  $I_{se} (\Omega)$
  can be evaluated exactly at small $\Omega$. We  obtain $I_{se} (\Omega) = I_{se} (0) + 3 i {\bar U} m  \Omega/(2 \pi \Sigma^{''} (T))$, where $ \Sigma^{''} (T)$ is the imaginary part of the fermionic self-energy at zero frequency.
  If we stop here and
  associate $(4/3) I_{se} (\Omega)$ with the fully renormalized  $\chi (0,\Omega)$
    to order ${\bar U}$,
   we would conclude that $\Gamma (q=0)$  becomes finite, in apparent violation of the spin conservation.
  It turns out, however~\cite{cm_prl}, that there  two other
 contributions to $\chi (0,\Omega)$  to order ${\bar U}$.
 They come from the two
  AL diagrams~\cite{al} shown in Fig. 1c. By power counting, these
 diagrams are of order ${\bar U}^2$, but we
  will see
 that
 one power of ${\bar U}$ actually
 cancels out.

  The two AL diagrams are equivalent and add up~\cite{comm}, such that
  one can consider one of them and multiply the result by $2$.
 In explicit form we have~\cite{see}
   \begin{widetext}
   \bea
  I_{AL} (\Omega) = 16  \int \int \int && G(\mb{k}_1,\omega_1) G(\mb{k}_1, \omega_1-\Omega) G(\mb{k}_2, \omega_2) G(\mb{k}_2, \omega_2-\Omega) G(\mb{k}_1+\mb{q'}, \omega_1 + \Omega')
  G(\mb{k}_2+\mb{q'}, \omega_2 + \Omega') \nonumber \\
  && \times U_{eff}(\mb{q'},\Omega') U_{eff} (\mb{q'}, \Omega' + \Omega)
  \label{4}
  \eea
  where
  $\int \int \int \equiv \int \frac{d^dk_1 d\omega_1}{(2 \pi)^{d+1}} \int\frac{d^dk_2 d\omega_2}{(2 \pi)^{d+1}}  \int \frac{d^d q' d\Omega'}{(2 \pi)^{d+1}}$.
  We use the same identity
 for the Green's functions as before, but also express the product of the two effective interactions as
  \beq
  U_{eff} (\mb{q'},\Omega') U_{eff} (\mb{q'}, \Omega' + \Omega) =  \frac{U_{eff} (\mb{q'},\Omega') - U_{eff} (\mb{q'}, \Omega' + \Omega)}{U^{-1}_{eff} (\mb{q'}, \Omega' + \Omega)-
  U^{-1}_{eff} (\mb{q'}, \Omega')}
  = \frac{U_{eff}(\mb{q'},\Omega') - U_{eff} (\mb{q'}, \Omega' + \Omega)}{\chi(\mb{q'},\Omega') -
  \chi (\mb{q'}, \Omega' +\Omega)}
  \label{5}
  \eeq
   \end{widetext}
where, we recall, $U^{-1}_{eff} = 1/{\bar U} - \chi({\bf q}, \Omega)$ and $\chi (\mb{q'}, \Omega') = 2 \int G(\mb{k}, \omega) G(\mb{k}+\mb{q'}, \omega + \Omega')$.
  We see that the r.h.s. of Eq. (\ref{5}) is of order ${\bar U}$, not ${\bar U}^2$, as one could assume by looking at the l.h.s. of this equation.
 This cancellation of one power of ${\bar U}$  is the
natural consequence of the fact that the
dispersion of
the effective interaction $U_{eff} ({\bf q}, \Omega)$
is due to the interaction with the same fermions whose susceptibility we consider.

Substituting (\ref{5}) into (\ref{4}) we find after some algebra~\cite{see} that
 $I_{AL} (\Omega)$
has the same form as self-energy and MT contributions and is given by
  \be
  I_{AL} (\Omega)= -8
   A(\Omega)
  \label{6}
  \ee
As a result,  $ I_{tot} (\Omega) = I_{se} (\Omega) +  I_{MT} (\Omega) +  I_{AL} (\Omega) =0$, i.e., $\Gamma (q)$ vanishes at $q=0$, as it indeed should
 for consistency
 with the spin conservation principle.
For completeness, we analyzed other diagrams which contain ${\bar U}^2$ in the prefactor but found that they remain of order ${\bar U}^2$, up to logarithmic corrections. Only in the two AL diagrams one power of ${\bar U}$ is cancelled out.

{\it Two-component systems.}--- We now analyze how this result changes when we consider a two-component system consisting of localized
and itinerant electrons.
As we discussed in the
 Introduction, we focus on the range $v_F q  \sim \Omega$ where the full dynamical susceptibility almost coincides with the one
 for localized spins, $\chi_L ({\bf q}, \Omega)$. By itself (i.e., with no itinerant fermions present) $\chi_L (0,\Omega)$ vanishes. We consider
 how it is modified due to the interaction with the itinerant electrons.

We use the same  model as before
 with the effective spin-fermion coupling
\be
H =
\Phi \sum_{\mb{q}} {\bf s}_{\mb q}  \cdot \mb{S}_\mb{-q}
\label{6_1}
\ee
\noindent
where  ${\bf s}_{\mb q}  = \sum_{\mb{k},\alpha\beta} c^\dagger_{{\mb k},\alpha} {\bf \sigma}_{\alpha\beta} c_{{\mb k} + {\mb q},\beta}$
   with $c^\dagger_{\mb{k} \alpha}, c_{\mb{k} \alpha}$ being the creation and annihilation operators for itinerant electrons, ${\mb \sigma}_{\alpha, \beta}$ are Pauli matrices, and
   $\mb{S}_\mb{q}$ describes the localized spins.  This interaction gives rise to the correction to the susceptibility of the localized spins
 $\chi^{tot}_{L} (0, \Omega) = \chi_L (0,\Omega) + I_{tot} (\Omega)$, where $I_{tot} (q,\Omega)$ is the fully renormalized
  spin-polarization bubble of itinerant fermions.
   Simultaneously,
   Eq.(\ref{6_1}), taken to second order, gives rise to effective interaction between itinerant carriers, mediated by the localized spins: $\Phi_{eff} (q, \Omega) \sim \Phi^2 \chi_L (q,\Omega)$, like in Fig. \ref{fig1}(a-c). The crucial difference with the previous case is that now the localized spins have their own dynamics even in the absence of the interaction with itinerant carriers.  This dynamics is  consistent with the conservation of the total spin of localized carriers, e.g.,  in a paramagnetic state
     it is
  spin diffusion: $\chi_L (q, \Omega) = \chi_0/(q^2 + \xi^{-2} -i \Omega/(Dq^2))$.
  However, when we include the spin-spin interaction between localized and itinerant carriers, we find that the
 contributions from AL diagrams no longer cancel out the contribution from self-energy and MT terms because the difference $\chi^{-1}_L (\mb{q'}, \Omega')-
\chi^{-1}_L (\mb{q'}, \Omega'+ \Omega)$ in the analogue of Eq (\ref{5}) is non-zero even when $\Phi=0$.
 As the consequence, the extra $\Phi^2$ in the AL diagrams does not cancel out, and the AL contribution $I_{AL} (\Omega)$
   becomes
    small compared to self-energy and MT terms. The sum of these two is, according to Eq. (\ref{2}), $(4/3) I_{se} (\Omega) = A + i B \Omega/\Sigma^{''} (T)$, where $A$ and $B$ are constants.
 Then, to leading order in $\Phi$,
   \be
   \chi^{tot}_{L}  (q \to 0,\Omega)=
   \frac{4}{3} I_{se} (\Omega) = A +i B \frac{\Omega}{\Sigma^{''} (T)}
   \label{7}
   \ee
   i.e., the system has a non-zero  $\Gamma_L (q \to 0) \propto \Sigma^{''} (T)$.
   This result holds for both 2D and 3D systems.

At  small but finite $q$ and at $\Omega < Dq^2$  the static part of $\chi_L (0,\Omega)$ can be approximated by $\chi_0\xi^2$, and once the damping due to
 interaction with itinerant carriers exceeds the diffusion term, the susceptibility and the structure factor  become
    \beq
     \chi^{tot}_{L}  ({\bf q},\Omega) \approx  \frac{\chi_0 \xi^2}{1 - i \frac{\Omega}{\Gamma_L (q)}};~ S({\bf q}, \Omega) \propto \frac{\chi_0 \xi^2 \Gamma_L (q)}{\Omega^2 +  \Gamma^2_L (q)}
   \label{7_1}
   \ee
   where $\Gamma_L (q)  \approx \Gamma_L (q=0) \propto \Phi^2 \Sigma^{''} (T) \xi^{-2}$.

The physical reason for a non-zero $\Gamma_L (q=0)$ is the fact that, when itinerant and localized electrons interact via spin-spin coupling ${\mb S} \cdot {\mb s}$,
   only the total
 combined spin is conserved, while the localized spin ${\mb S}$
can change its $z-$component and transfer the difference to the spin ${\bf s}$ of an itinerant electron. This reasoning parallels the one presented by Mineev~\cite{Mineev2013}. He also found $\Gamma_L (q=0)
 \propto \xi^{-2}$.
Our microscopic consideration expresses $\Gamma_L (q=0) $ in terms of the coupling $\Phi$ and the temperature dependent fermionic self-energy of the itinerant electrons $\Sigma^{''} (T)$.  This allows us to proceed further with the analysis of the temperature dependence of $\Gamma (0)$ and the comparison with the experimental data.

\begin{figure}[t]
\begin{center}
\includegraphics[scale=0.60]{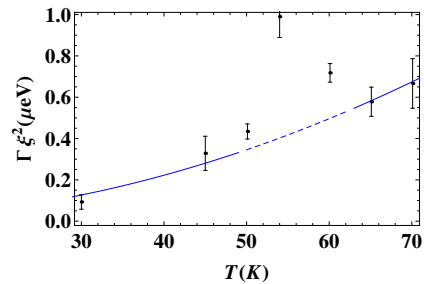}
\end{center}
\vspace{-0.5cm}
\caption{\label{fig:Fig2} The fitting of the data for $\Gamma (T) \xi^2$ from the work of Ref.~\onlinecite{Huxley2003}. The solid line is the fit by the dependence $\propto T^2$.
The dashed portion of the line shows the regime around $T_c \simeq 54 K$ where $\Gamma \xi^2 \propto \Sigma^{''} (T)$ gets an additional enhancement due to the increase of $\Sigma^{''} (T)$.}
\label{fig:2}
\end{figure}

{\it Fermionic self-energy.}---The diagram for the fermionic self-energy is presented in Fig. 1d. The wavy line is the propagator of localized spins $\chi_L$ which we assume to have a diffusive form.
Using the spectral representation of $G(\mb{k}, \omega)$ and $\chi_L (\mb{k}+\mb{q}, \omega)$, we obtain
\be
 \Sigma^{''} (T) \propto  \Phi^2 \int \frac{d\Omega \Omega}{\sinh{\frac{\Omega}{T}}} \int d^{d-1} q\frac{D q^2}{D^2 q^4 (\xi^{-2} + q^2)^2 + \Omega^2}
\label{8}
\ee
 At some distance away from the transition point,  $\Sigma^{''} (T) \propto T^{3/2}$ in $2D$ and $(T^2/E_F) \log{E_F/T}$ in $3D$.
This dependence holds both in the paramagnetic phase (where Eq. (\ref{8}) is valid), and in the ferromagnetic state. Right at the transition point, $T=T_c$,  $\xi= \infty$,
$\Sigma^{''} (T)$ is enhanced: a self-consistent solution yields
$\Sigma^{''} (T) \propto T \log T$  in 3D and $\Sigma^{''} (T) \propto T^{4/5}$ in 2D.
As the result, $\Gamma_L (T) \xi^2 \propto \Sigma^{''} (T)$
scales as  some power of $T$  below and above the transition, but get an enhancement very near the transition point.
 In  Fig.\ref{fig:2} we compare our theory
 with the
 experimental data for $\Gamma ({\bf q}, T) \xi^2$ as a function of $T$
 in UGe$_2$
 (Ref.\onlinecite{Huxley2003}).
We recall that we identify $\Gamma_L ({\bf q},T)$ with the measured $\Gamma ({\bf q}, T)$ over the range $v_F q \geq \Omega$ probed by the experiments~\cite{comm_new}.
At asymptotically small  $v_F q \ll \Omega$, $\Gamma ({\bf q}, T)$ should indeed vanish.
The data are consistent with our
  result that there is a smooth increase of $\Gamma (T) \xi^2$ with $T$ in both the ferromagnetic and paramagnetic state, on top of which there is a peak at $T_c$.

{\it Conclusion.}---To conclude, in this paper we presented the microscopic study of the damping term of spin excitations in a metal near a ferromagnetic instability. We demonstrated that in an one-component systems, the spin scattering rate $\Gamma (q)$ vanishes at $q=0$ as the consequence of the spin conservation. We argued that to see this in a loop expansion, one needs to invoke AL scattering processes.
 We then considered a two-component model with localized and itinerant fermions, which was argued~\cite{Yaouanc2002, Mineev2013}
to describe ferromagnetic uranium compounds such as UGe$_2$ and UCoGe.
 Localized spins mostly contribute to long-range order and to Curie susceptibility and for $v_F q \geq \Omega$, probed by the experiments, the measured damping rate
 $\Gamma ({\bf q},T)$ almost coincides with $\Gamma_L ({\bf q},T)$  for localized spins.
We showed that
   for $\Gamma_L ({\bf q},T)$
  the AL diagrams
   are relatively small. Without the AL contribution, the spin scattering rate $\Gamma_L (q=0)$ becomes finite and scales with the temperature dependent part of fermionic self-energy  $\Sigma^{''} (T)$ for itinerant fermions.
   We
   found that
$\Gamma_L (0,T) \xi^2$ has a peak at $T_c$.
This
 is consistent with the
 data on UGe$_2$.

We acknowledge helpful discussions with Andrew Huxley, Vladimir Mineev, and Dmitrii Maslov. We also thank A.Huxley for providing us the precise experimental data for the Fig. 2. The
work was supported by the DOE grant DE-FG02-ER46900 and by a Leverhulme Visiting Professorship held at the University of Loughborough
(AVC). JJB was supported by the EPSRC grant EP/H049797/1.

\end{document}


\section{Supplemental Material.}



\subsection{Cancellation of the  diagrams of the spin polarization bubble at zero momentum and a finite frequency}

\subsubsection{A. Charge channel}
We show how self-energy and Maki-Thompson (MT) vertex correction insertions into the particle-hole bubble   cancel each other in the charge channel.
The diagrams are shown in Fig. 1 in the text. For charge channel, side vertices are spin delta-functions rather than $\sigma-$matrices. 
For convenience, we perform calculations in Matsubara frequencies. 

The expressions for the two  self-energy diagrams in Fig. 1b are:
\begin{eqnarray}
I_{se}^A (\Omega)  = 3 \int \frac{d^dq d\Omega'}{(2 \pi)^{d+1}} G^2({\bf{k}}, \omega) G({\bf{k}}+{\bf{q}}, \omega+\Omega') G({\bf{k}}+{\bf{q}}, \omega-\Omega) U_{eff}({\bf{q}}, \Omega')
\end{eqnarray}

\begin{eqnarray}
I_{se}^B (\Omega) = 3 \int \frac{d^dq d\Omega'}{(2 \pi)^{d+1}} G^2({\bf{k}}, \omega) G({\bf{k}}+{\bf{q}}, \omega+\Omega') G({\bf{k}}+{\bf{q}}, \omega+\Omega) U_{eff}({\bf{q}}, \Omega')
\end{eqnarray}
\noindent where the factor of $3$ is due to summation over spin indices, the Green's functions are
 $G(\vec{k}, \omega) = \frac{1}{i \omega - \epsilon({\bf{k}})} $, and the expression for $U_{eff} ({\bf{q}}, \Omega)$ in given in the main text:  $U_{eff} ({\bf{q}}, \omega) \approx {\bar U}/(1 -{\bar U} \chi ({\bf q},\Omega)))$.  The $\chi({\bf q},\Omega)$ in this expression is the fully renormalized spin-polarization bubble, but to lowest order in the coupling ${\bar U}$ it can be replaced by the spin susceptibility of free fermions.

By using several times the known identity:
\begin{equation}
G_1 G_2 = (G_1 - G_2) \times \frac{1}{G_2^{-1} - G_1^{-1}}
\label{y1}
\end{equation}

\noindent we obtain:

\begin{eqnarray}
\nonumber
I_{se}^A (\Omega) = && 3 \int \frac{d^dq d\Omega'}{(2 \pi)^{d+1}} G({\bf{k}}+{\bf{q}}, \omega+\Omega') G({\bf{k}}, \omega) U_{eff}({\bf{q}}, \Omega')  \left[G({\bf{k}}, \omega) - G({\bf{k}}, \omega-\Omega)\right] \frac{1}{(-i \Omega)} \\
\nonumber
=&& -\frac{3}{(i \Omega)} \int \frac{d^dq d\Omega'}{(2 \pi)^{d+1}}  G^2({\bf{k}}, \omega) G({\bf{k}}+{\bf{q}}, \omega+\Omega') U_{eff}(\vec{q}, \Omega') \\
&&+ \frac{3}{{\Omega}^2} \int \frac{d^dq d\Omega'}{(2 \pi)^{d+1}} G({\bf{k}}+{\bf{q}}, \omega+\Omega') U_{eff}({\bf{q}}, \Omega') \left [G({\bf{k}}, \omega) - G({\bf{k}}, \omega-\Omega)\right]
\end{eqnarray}

\begin{eqnarray}
\nonumber
I_{se}^B (\Omega) =&& 3 \int \frac{d^dq d\Omega'}{(2 \pi)^{d+1}} G({\bf{k}}+{\bf{q}}, \omega+\Omega') G({\bf{k}}, \omega) U_{eff}({\bf{q}}, \Omega')  \left[G({\bf{k}}, \omega) - G({\bf{k}}, \omega+\Omega) \right] \frac{1}{(-i \Omega)} \\
\nonumber
= &&\frac{3}{(i \Omega)} \int \frac{d^dq d\Omega'}{(2 \pi)^{d+1}}  G^2({\bf{k}}, \omega) G({\bf{k}}+{\bf{q}}, \omega+\Omega') U_{eff}({\bf{q}}, \Omega') \\
&& + \frac{3}{{\Omega}^2} \int \frac{d^dq d\Omega'}{(2 \pi)^{d+1}} G({\bf{k}}+{\bf{q}}, \omega+\Omega') U_{eff}({\bf{q}}, \Omega') \left[G({\bf{k}}, \omega) - G({\bf{k}}, \omega+\Omega)\right]
\end{eqnarray}

Therefore the total contribution  from the two diagrams is:

\begin{eqnarray}
I_{se} (\Omega) = I_{se}^A (\Omega) + I_{se}^B (\Omega) =  \frac{3}{\Omega^2} \int \frac{d^dq d\Omega'}{(2 \pi)^{d+1}} \left[2G({\bf{k}}, \omega) -  G({\bf{k}}, \omega+\Omega)- G({\bf{k}}, \omega-\Omega) \right]  G({\bf{k}}+{\bf{q}}, \omega+\Omega') U_{eff}({\bf{q}}, \Omega')
\end{eqnarray}

Similarly, the vertex-correction MT diagram reads:
$$
I_{MT}= 3 \int \frac{d^dq d\Omega'}{(2 \pi)^{d+1}} G({\bf{k}}, \omega) G({\bf{k}}+{\bf{q}}, \omega+\Omega') G({\bf{k}}+{\bf{q}}, \omega-\Omega +\Omega') U_{eff}({\bf{q}}, \Omega')
$$
Using (\ref{y1}) we rewrite it as
\begin{eqnarray}
\nonumber
I_{MT} (\Omega) = &&-\frac{3}{\Omega^2} \int \frac{d^dq d\Omega'}{(2 \pi)^{d+1}} \left[G(\mb{k}, \omega) - G(\mb{k}, \omega-\Omega)\right] \left[G(\mb{k}+{\bf{q}}, \omega+\Omega') - G(\mb{k}+{\bf{q}}, \omega-\Omega+\Omega')\right] U_{eff}({\bf{q}}, \Omega') \\
\nonumber
= && -\frac{3}{\Omega^2} \int \frac{d^dq d\Omega'}{(2 \pi)^{d+1}} \left[G(\mb{k}, \omega) - G(\mb{k}, \omega-\Omega)\right] G(\mb{k}+{\bf{q}}, \omega+\Omega')  U_{eff}({\bf{q}}, \Omega') \\
&& + \frac{3}{\Omega^2} \int \frac{d^dq d\Omega'}{(2 \pi)^{d+1}} \left[G(\mb{k}, \omega) - G(\mb{k}, \omega-\Omega)\right] G(\mb{k}+{\bf{q}}, \omega-\Omega+\Omega') U_{eff}({\bf{q}}, \Omega')
\end{eqnarray}

Shifting the frequency: $\omega'=\omega-\Omega$ in the last line and then adding the two terms, we finally obtain:
\beq
I_{MT} = -\frac{3}{\Omega^2} \int \frac{d^dq d\Omega'}{(2 \pi)^{d+1}} \left[ 2G(\mb{k}, \omega) -  G(\vec{k}, \omega+\Omega)- G(\mb{k}, \omega-\Omega) \right]  G(\mb{k}+{\bf{q}}, \omega+\Omega') U_{eff}({\bf{q}}, \Omega')
\label{y3}
\eeq

Adding $I_{se}$ and $I_{MT}$ together, we obtain in the charge channel:
\begin{equation}
I_{se} (\Omega) + I_{MT} (\Omega) = 0
\end{equation}
There are also two Aslamazov-Larkin (AL) diagrams (Fig. 1c). They are identical, up to sign. For charge channel (side vertices with spin $\delta-$functions) signs are different, and the  the two AL diagram cancel each other. Then the total $I_{se}  (\Omega) + I_{MT} (\Omega) + I_{AL} (\Omega) =0$, as it is required by charge conservation.

\subsubsection{B. Spin Channel}
For spin channel, the side vertices contain spin $\sigma-$matrices ${\bf \sigma}_{\alpha\beta}$.
The overall sign in $I_{se}$ is now $6$, while the overall sign in $I_{MT}$ is $(-2)$.
Then $I_{MT} (\Omega) = 1/3 I_{se} (\Omega)$ and $I_{se} (\Omega) + I_{MT} (\Omega) = (4/3) I_{se} (\Omega)$.  In explicit form
\beq
\frac{4}{3} I_{se} = -\frac{8}{\Omega^2} \int \frac{d^dq d\Omega'}{(2 \pi)^{d+1}} \left[ 2G(\mb{k}, \omega) -  G(\mb{k}, \omega+\Omega)- G(\mb{k}, \omega-\Omega) \right]  G(\mb{k}+\mb{q}, \omega+\Omega') U_{eff}(\mb{q}, \Omega')
\label{y4}
\eeq

Simultaneously, however, the two AL diagrams add up instead of canceling out. The total contribution from the two AL diagrams is:

\begin{eqnarray}
\nonumber
I_{AL} (\Omega) = 16 \int \frac{d^dk d\omega} {(2 \pi)^{d+1}} \int \frac{d^2q d\Omega'}{(2 \pi)^3} \int \frac{d^dp d\omega'}{(2 \pi)^{d+1}}  G(\mb{k}, \omega) G(\mb{k}, \omega - \Omega) G({\bf{p}}, \omega') G(\mb{k}+{\bf{q}}, \omega+\Omega') \\
\times \left[ G({\bf{p}}, \omega'+\Omega) G({\bf{p}}-{\bf{q}}, \omega'-\Omega') + G({\bf{p}}, \omega'-\Omega) G({\bf{p}}+{\bf{q}}, \omega+ \omega'+ \Omega') \right]  U_{eff}({\bf{q}}, \Omega')  U_{eff}({\bf{q}}, \Omega+\Omega')
\end{eqnarray}

%

The product of the two effective interactions can be rewritten in the form:
\begin{eqnarray}
  U_{eff} (\mb{q'},\Omega') U_{eff} (\mb{q'}, \Omega' + \Omega) &=&  \frac{U_{eff} (\mb{q'},\Omega') - U_{eff} (\mb{q'}, \Omega' + \Omega)}{U^{-1}_{eff} (\mb{q'}, \Omega' + \Omega)-
  U^{-1}_{eff} (\mb{q'}, \Omega')}
  \nonumber \\
  && = \frac{U_{eff}(\mb{q'},\Omega') - U_{eff} (\mb{q'}, \Omega' + \Omega)}{\chi(\mb{q'},\Omega'+\Omega) -
  \chi (\mb{q'}, \Omega')}
  \label{y2}
  \end{eqnarray}
where, we recall, $\chi (\mb{q'}, \Omega') = 2 \int G(\mb{k}, \omega) G(\mb{k}+\mb{q'}, \omega + \Omega')$. Note that the resulting expression is of first order in $U{eff}$.  Physically, this is the consequence of the fact the dynamics of collective excitations comes solely from the interaction with fermions (i.e., that there is
 no damping of a collective boson if ${\bar U} =0$).

  Using (\ref{y2}) for $ U_{eff} (\mb{q'},\Omega') U_{eff} (\mb{q'}, \Omega' + \Omega)$ and
  using the identity Eq.~(3) several times, we obtain:
\begin{eqnarray}
\nonumber
I_{AL} (\Omega) = 16  \int \int \int G(\mb{k}, \omega) G(\mb{k}, \omega - \Omega) G(\mb{p}, \omega') G(\mb{p}, \omega'-\Omega) G(\mb{k}+\mb{q}, \omega+\Omega') G(\mb{p}+\mb{q}, \omega'+\Omega') \\
\nonumber
\times \left[U_{eff}(\mb{q}, \Omega') - U_{eff}(\mb{q}, \Omega'+\Omega) \right] \frac{1}{2 u \left[\Pi(\mb{q}, \Omega' +\Omega)- \Pi(\mb{q}, \Omega') \right] } \\
\nonumber
= -\frac{16}{\Omega^2} \int \int \int \left[ G(\mb{k}, \omega) - G(\mb{k}, \omega - \Omega) \right] \left[ G(\mb{p}, \omega') - G(\mb{p}, \omega'- \Omega) \right] G(\mb{k}+\mb{q}, \omega+\Omega') G(\mb{p}+\mb{q}, \omega'+\Omega') \\
\nonumber
\times \left[ U_{eff}(\mb{q}, \Omega') - U_{eff}(\mb{q}, \Omega'+\Omega) \right]  \frac{1}{2 u \left[\Pi(\mb{q}, \Omega' +\Omega)- \Pi(\mb{q}, \Omega') \right] } \\
\nonumber
-\frac{16}{\Omega^2} \int \int \int \left[ G(\mb{k}, \omega) G(\mb{p}, \omega')  - G(\mb{k}, \omega - \Omega) G(\mb{p}, \omega') - G(\mb{k}, \omega) G(\mb{p}, \omega'-\Omega) + G(\mb{k}, \omega - \Omega) G(\mb{p}, \omega'-\Omega) \right] \\
\nonumber
 \times G(\mb{k}+\mb{q}, \omega+\Omega') G(\mb{p}+\mb{q}, \omega'+\Omega') \left[ U_{eff}(\mb{q}, \Omega') - U_{eff}(\mb{q}, \Omega'+\Omega) \right] \frac{1}{2 u \left[\Pi(\mb{q}, \Omega' +\Omega)- \Pi(\mb{q}, \Omega') \right] }
\end{eqnarray}
 where we used short-hand notation $\int \int \int \equiv \int \frac{d^dk d\omega}{(2 \pi)^{d+1}}\int \frac{d^dq d\Omega'}{(2 \pi)^{d+1}} \int \frac{d^dp d\omega'}{(2 \pi)^{d+1}}$.

At this stage we can separate the integration $\int \frac{d^dp d\omega'}{(2 \pi)^{d+1}}$ of a product of two Green's functions in each term and rewrite the last equation as
\begin{eqnarray}
\nonumber
I_{AL} (\Omega)  =&& -\frac{8}{\Omega^2} \int \int \frac{ \left[ U_{eff}(\mb{q}, \Omega') - U_{eff}(\mb{q}, \Omega'+\Omega) \right]} {2 u \left[ \Pi(\mb{q}, \Omega' +\Omega)- \Pi(\mb{q}, \Omega') \right] } [  G(\mb{k}, \omega) G(\mb{k}+\mb{q}, \omega+\Omega') \Pi(\mb{q}, \Omega')  + G(\mb{k}, \omega-\Omega) G(\mb{k}+\mb{q}, \omega+\Omega') \\
\nonumber
&& \times \Pi(\mb{q}, \Omega' + \Omega) - G(\mb{k}, \omega-\Omega) G(\mb{k}+\mb{q}, \omega+\Omega') \Pi(\mb{q}, \Omega') - G(\mb{k}, \omega) G(\mb{k}+\mb{q}, \omega+\Omega') \Pi(\mb{q}, \Omega' + \Omega) ] \\
\nonumber
=&&\frac{8}{\Omega^2} \int \int \left[ U_{eff}(\mb{q}, \Omega') - U_{eff}(\mb{q}, \Omega'+\Omega) \right] \left[ G(\mb{k}, \omega) G(\mb{k}+\mb{q}, \omega+\Omega') - G(\mb{k}, \omega-\Omega) G(\mb{k}+\mb{q}, \omega+\Omega') \right] \\
=&&\frac{8 u^2}{\Omega^2} \int \int  U_{eff}(\mb{q}, \Omega') G(\mb{k}+\mb{q}, \omega+\Omega') \left[ 2 G(\mb{k}, \omega) - G(\mb{k}+\mb{q}, \omega+\Omega') - G(\mb{k}, \omega-\Omega') \right]
\end{eqnarray}
\noindent where now  $\int \int \equiv \int \frac{d^dk d\omega}{(2 \pi)^{d+1}}\int \frac{d^dq d\Omega'}{(2 \pi)^{d+1}} $, and the last
 expression is obtained by shifting the frequency variable $\Omega'+\Omega$ to $\Omega'$ in $U_{eff}(\mb{q}, \Omega'+\Omega)$ and in the corresponding Green's functions.
Comparing this expression with Eq. (\ref{y4}) we immediately see that 
\beq
I_{AL} (\Omega)= -\frac{4}{3} I_{se} (\Omega),
\eeq 
such that  
the total contribution to the spin polarization bubble  
\beq
I_{se} (\Omega) +I_{MK} (\Omega) +I_{AL} (\Omega) = 0,
\eeq
 as it should be due to spin conservation.